\documentclass[a4paper]{article}

\usepackage{array,graphicx}
\usepackage{amssymb,color,amsmath}
\usepackage[pdftex]{hyperref}
\usepackage[round,colon]{natbib}
\usepackage[normalem]{ulem}
\usepackage{authblk}

\newcommand{\myfigurewidth}{25pc}

\defcitealias{AbrSte1965}{A\&S}

\title{{Analytical model of infiltration under constant surface ponding}}
\author[1,2]{Dimetre Triadis}
\author[1]{Philip Broadbridge}
\affil[1]{Department of Mathematics and Statistics, La Trobe University, Victoria, Australia}
\affil[2]{Institute of Mathematics for Industry, Kyushu University, Japan}
\date{}
\begin{document}
\maketitle

\begin{abstract}
An analytical solution of the nonlinear Richards equation is presented, for one-dimensional infiltration into a soil of uniform initial moisture content subject to a constant depth of surface ponded water. Adopted mathematical forms of the soil water diffusivity and conductivity are flexible enough to model a range of real soils. The solution takes the form of a power series in $\sqrt t$, but is observed to converge not only for small times but also for relatively large times at which travelling-wave-like behavior is evident. The solution is used to tabulate exact infiltration coefficients with higher-order corrections as the natural nonlinear limit of soil properties is approached. Previously published approximate solutions that apply for a wide range of soil properties are tested against the exact solution and found to be sufficiently accurate.  
\end{abstract}

%
%

\section{Introduction}
Infiltration under constant-pressure-head boundary conditions is important for a number of reasons. Firstly, it is a simple representation of soil-water flow under a surging river or a flooding irrigation system applied to a river bed or irrigation furrow comprised of soil at initial moisture content below saturation level. Secondly the hydrological behaviour under that simply expressed canonical boundary condition aids our general conceptual foundations of ponded infiltration that allows us to consider more general situations. Thirdly, an understanding of those hydrodynamics enables us to apply inverse problems to infer the numerical values of key soil hydraulic parameters \citep{AlakaylehFangClement2019,RegaladoRitterAlvarez-BenediMunoz-Carpena2005,AhmedNestingenNieberGulliverHozalski2014}. These ideal boundary conditions with uniform initial water content can be well approximated in soil laboratories \citep{BondCollis-George1981}. \\

Mathematical modelling of ponded infiltration must simultaneously confront two major difficulties. The first is that the positive pressure sets up a saturated zone which extends downwards to an unknown location of a free boundary where an unsaturated zone begins. The second is that the Darcy-Buckingham formulation of water transport in unsaturated soil, involves the highly nonlinear Richards equation, a diffusion-convection equation in which both the diffusivity $D$ and the `velocity coefficient'  $K'(\theta)$ expressed in terms of hydraulic conductivity, depend strongly on the volumetric water content $\theta$, which is the dependent variable of the mathematical problem to be solved. In many applications, most of the details of the functions $D(\theta)$ and $K(\theta)$ are ignored, in favour of the original 1911 Green--Ampt model \citet{GreenAmpt1911} that assumes a plug flow with a step function water content profile, penetrating the soil under the potential gradient between the surface positive pressure head and a constant negative `suction head' $H$ at the wetting front. In Buckingham's formulation, the latter is more correctly expressed as a constant negative potential energy of water at the wetting front \citep{Buckingham1907}. Within the general theory of unsaturated soil-water, after assuming a {\it constant} wetting front potential, the Richards equation reduces to the Green--Ampt model under the assumption of a Dirac delta function diffusivity $D=\delta(\theta-\theta_s)$. The Green--Ampt model provides a first approximation to the cumulative infiltration function $i(t)$ which is the equivalent depth of free water having entered the soil. Philip (1992) pointed out that solutions of the Green--Ampt model under falling-head boundary conditions, take the same form as for constant-head; only the values of the constants change. \citet{WarrickZerihunSanchezFurman2005} showed that this simple model could similarly be solved under specified time-dependent pressure head at the top surface. \citet{KacimovAlIsmailyAlMaktoumi2010} extended the Green--Ampt model to heterogeneous soils.\\

The delta-function diffusivity that underlies the Green--Ampt model, requires that $D(\theta) =0$ for $\theta<\theta_s$. Furthermore, it accounts for only two values of the hydraulic conductivity, namely  $K_s$ at saturation and $K_n$ at the initial moisture content $\theta_n$. As pointed out by \citet{BarryParlangeHaverkamp1995}, the nature of conductivity function $K(\theta)$ near saturation does make a difference to the flow even with a delta-function diffusivity.\nocite{Philip1992a}  In fact physically reasonable behaviour in the delta-function diffusivity limit depends on how $K(\theta)$ approaches a step function in the limit as $D(\theta)$ approaches a delta function \citep{TriadisBroadbridge2012}. Taking the limiting form of the integrable soil hydraulic model of \citet{BroadbridgeWhite1988a} that gives a delta-function diffusivity, the expression for time to ponding under constant irrigation rate \citep{BroadbridgeWhite1987} is exactly that of \citet{SmithParlange1978} which is much more accurate than that given by the Green--Ampt model. Furthermore, under constant-concentration boundary conditions \citep{TriadisBroadbridge2010}, that limiting form gives an infiltration series 
$$i(t)=St^{1/2}+\frac{1}{3}(K_s-K_n) t+K_n t+O(t^{3/2})$$
which agrees much better with experiment \citep{Talsma1969} than that of the Green--Ampt model which has 
$$i(t)=St^{1/2}+\frac{2}{3}(K_s-K_n)t+K_n t+ O(t^{3/2}).$$
 This paper concentrates on extending these results to the case of ponded infiltration. With antecedents back to \citet{TalsmaParlange1972} and \citet{ParlangeLisleBraddockSmith1982}, \citet{ParlangeHaverkampTouma1985} and \citet{HaverkampParlangeStarrSchmitzFuentes1990} have developed widely applicable approximate relationships between time and infiltration rate, that have extra parameters to account for some of the structure lost in the delta-function-diffusivity limit, see also \citet{BarryParlangeHaverkampRoss1995}. Using the finite element method, \citet{Mollerup2007} solved a realistic van Genuchten soil model under experimental boundary conditions of variable pressure head. At all times the numerical infiltration rate agreed very well with the expression of \citet{ParlangeHaverkampTouma1985}. The calculated 6-term Philip Infiltration series was calculated and shown to agree extremely well up to the order of a gravity time scale $t_{grav}$ after which the calculated series became inappropriate.\\

Flow in unsaturated soil under constant-head boundary conditions  has not previously been solved analytically for any reasonably realistic soil hydraulic model. That is the main aim of this current work. The most general integrable model was previously solved in \citet{TriadisBroadbridge2010} with constant-concentration boundary conditions. Even in that case, the mathematical boundary value problem transforms to a difficult free boundary problem under the transformations that linearise the governing equation. In the case of constant-head infiltration, there is an additional physical free boundary. Nevertheless we are able to construct an exact solution as a series in $\sqrt t$. Using a large number of series terms, the solution is evaluated up to dimensionless times $t_*=5$, a time at which travelling wave behaviour is apparent. Infiltration coefficients are constructed in terms of standard soil parameters, plus two additional parameters of the integrable model. These are a nonlinearity parameter $C$ and a form factor $\zeta$ for the hydraulic conductivity. The original Green--Ampt model is recovered by taking the limit as $(C,\zeta)\to(1^+,0^+)$. Contrary to popular belief, the Green--Ampt model follows not from a step-function conductivity but from an unrealistic linear conductivity $K(\theta)$ that is compatible with the assumption of a fixed suction head at the wetting front. A much more realistic delta-function soil follows from $(C,\zeta)\to (1,1)$ that leads to an infiltration function resembling that of \citet{TalsmaParlange1972}  equation (7), in the limit as the ponded depth drops to zero.
We extend this result to constant-head ponded infiltration for which we provide an exact solution. This enables us to quantify the level of accuracy in the approximate infiltration functions of \citet{ParlangeHaverkampTouma1985} and \citet{HaverkampParlangeStarrSchmitzFuentes1990}.

\section{Transformation of the Richards equation}

\begin{figure}[h]
\begin{center}
\includegraphics[width=\myfigurewidth]{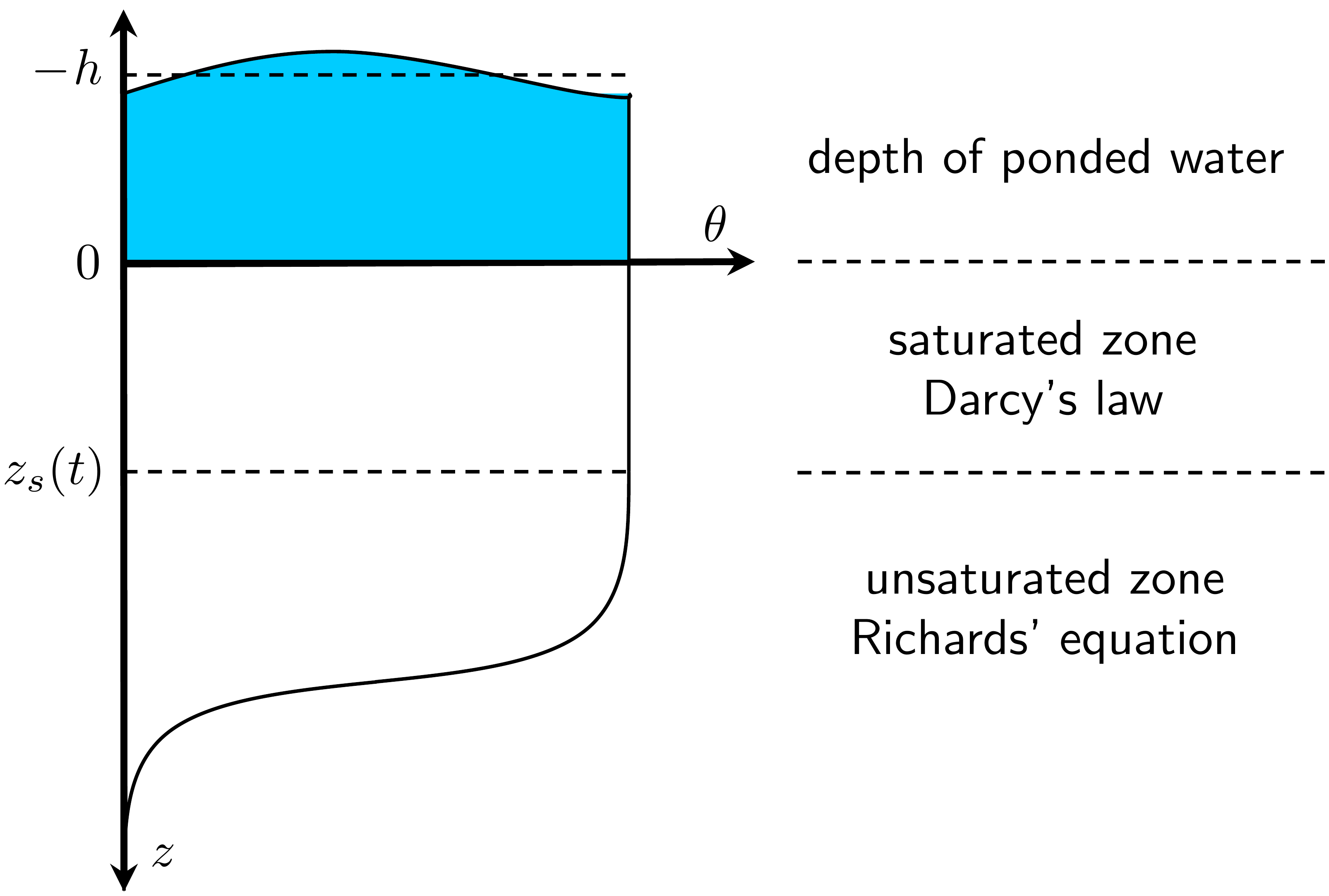}
\caption{Formation of a growing saturated zone under surface ponding}
\label{pondpic}
\end{center}
\end{figure}

We consider a one-dimensional model of infiltration of water into a soil whose volumetric moisture content $\theta$ is initially constant and unsaturated, equal to $\theta_n$. At time $t=0$, a depth of ponded water is placed on the soil surface, so that infiltration is initiated while the ponded depth is maintained at a constant height $h$. It is convenient to measure depth $z$ as positive-downward. As time progresses, a fully saturated zone with $\theta = \theta_s$ is formed below the soil surface, above an unsaturated region. A schematic illustration of the situation is shown in Fig.~\ref{pondpic}. Flow in the saturated region at depths less than $z_s$ is governed simply by Darcy's law, whereas the unsaturated flow at greater depths can be modelled using Richards' equation.  

Denoting the cumulative infiltration $i(t)$, the infiltration rate in the saturated zone is determined by the saturated hydraulic conductivity $K(\theta_s) = K_s$, and the moisture potential gradient:
\begin{equation}
i'(t) = K_s \left( 1 + \frac{h}{z_s(t)} \right). 
\label{Darcyraw}
\end{equation}
Assuming incompressibility, we can equate the flux $J(z,t)$ at the surface and the saturated front: $J(0, t) = J\big(z_s(t), t \big)$, so that total infiltration can be written in terms of the flux at the saturated front
\begin{equation}
i(t) = \int_0^t J\big(z_s(\tau), \tau \big)\, d\tau.
\label{fluxatfrontraw}
\end{equation}

For $z_s(t) < z < \infty$ we assume that infiltration is governed by the nonlinear Richards equation for $\theta(z,t)$
\begin{align}
\frac {\partial \theta}{\partial t} = \frac {\partial}{\partial z} \left[ D(\theta) \frac {\partial \theta}{\partial z} - K(\theta) \right], 
\end{align}
where $D(\theta)$ is the soil-water diffusivity, and $K(\theta)$ the hydraulic conductivity, with $K(\theta_n) = K_n$. We have an initial condition $\theta(z, 0) = \theta_n$ and the boundary condition $\theta\big(z_s(t)\big) = \theta_s$. 

By adopting the integrable soil model of \citet{BroadbridgeWhite1988a} we will be able to produce an analytical solution for $i(t)$ and $\theta(z, t)$. Hence we adopt the particular diffusivity and conductivity functions
\begin{align}  
D(\theta) &= \frac{a}{(b-\theta)^2}, \quad 
K(\theta) = \frac{\lambda}{2(b-\theta)} + \gamma(b-\theta) +\eta.
\label{intmodelDK}
\end{align}
As shown in \citet{TriadisBroadbridge2010} the above forms are versatile enough to satisfactorily model real soils. To reduce the number of parameters in our model we will adopt dimensionless variables $\Theta$, $z_*$ and $t_*$:
\begin{align}
&\begin{array}{ll}
\displaystyle
\Delta \theta = \theta_s - \theta_n,
& \displaystyle
\Delta K = K_s - K_n
\\
\displaystyle
l_s = \frac 1 {\Delta K} \int_{\theta_n}^{\theta_s} D(\theta) d\theta = \frac {\mathfrak H S_0^2}{C(C-1) \Delta \theta\, \Delta K},
& \displaystyle
t_s = \frac {\Delta \theta}{\Delta K}\, l_s,
\\
\displaystyle  \vphantom{\frac 11}
z = l_s z_*,
& \displaystyle 
t = t_s t_*,
\\
\displaystyle \vphantom{\frac 11}
\theta = \Delta \theta \,\Theta + \theta_n,
& \displaystyle
h = l_s \Delta \theta\, h_*,
\\
\displaystyle  \vphantom{\frac 11}
D(\theta) = \frac {l_s^2}{t_s} D_*(\Theta),
& \displaystyle
K(\theta) = \Delta K\, K_*(\Theta) + K_n,
\\
\displaystyle  \vphantom{\frac 11}
i_-(t) = i(t) - K_n t =  l_s \Delta \theta\, i_*(t_*),
& \displaystyle
\kappa_n = \frac {K_n}{\Delta K},
\\
\displaystyle  \vphantom{\frac 11}
J(z,t) = K(\theta) - D(\theta) \frac {\partial \theta}{\partial z} = \Delta K\, J_*(z_*, t_*) + K_n,
& \displaystyle
h_+ = \Delta \theta(1 + \kappa_n) h_*,
\\
\displaystyle  \vphantom{\frac 11}
C = \frac {b - \theta_n}{\Delta \theta},
& \displaystyle
\zeta = 1 + \gamma \frac {\Delta \theta}{\Delta K}.
\end{array}
\label{dimvariables}
\end{align}
Our length scale is written in terms of the sorptivity $S_0$, obtained by maintaining saturation without ponding at the surface of a soil in the initial state $\theta = \theta_n$. We have $a = \mathfrak H S_0^2$, where the factor $\mathfrak H/[C(C-1)]$ varies between $1/2$ and $\pi/4$ for $1 \le C < \infty$, as discussed in \citet{BroadbridgeWhite1988a}.

Equation (\ref{Darcyraw}) showing Darcy's law in the saturated zone then becomes
\begin{equation}
 z_{*s}(t_*) = \frac {h_+}{i'_*(t_*) -1}.
\label{Darcy}
\end{equation}
Note that the initial condition implies $z_{*s}(0) =0$, which is compatible with an unbounded infiltration rate $i'_*(t_*)$ as $t_* \to 0$. The calculation of the infiltration according to the flux at the saturated front (\ref{fluxatfrontraw}), retains its form
\begin{equation}
i_*(t_*) =  \int_0^{t_*} J_*\big(z_{*s}(\tau), \tau \big)\, d\tau.
\label{fluxatfront}
\end{equation}
The soil properties of the integrable model (\ref{intmodelDK}) may be written in terms of the two dimensionless parameters $C$ and $\zeta$, whose ranges and effects on soil properties are discussed in \citet{TriadisBroadbridge2010}. Hence we arrive at an integrable form of the Richards equation to be solved for $\Theta(z_*,t_*)$ in the unsaturated zone. 
\begin{align}
\qquad &\frac {\partial \Theta}{\partial t_*} = 
\frac {\partial }{\partial z_*}
\left[ 
\frac {C(C-1)}{(C-\Theta)^2}
\frac {\partial \Theta}{\partial z_*} 
- \zeta \frac {C(C-1)}{C-\Theta}
- (\zeta-1) \Theta
\right];
\\
\hspace{0.0cm} &0 < t < \infty, \qquad z_{*s}(t_*) \le z_* < \infty, \qquad z_{*s}(0) = 0;
\nonumber \\
&\Theta\Big(z_{*s}(t_*),t_*\Big) = 1,\quad {\rm for} \quad t_* \in (0,\infty);
\nonumber \\
&\Theta(z_*,0) = 0, \quad {\rm for} \quad z_* \in (0,\infty).
\nonumber
\end{align}
The transformation of this equation to a linear partial differential equation is basically the same as that described in \citet{TriadisBroadbridge2010}, with only minor and natural changes to account for the new moving boundary condition at the saturated interface. We outline it briefly below. 

Adopting a new spatial variable $Z = \zeta z_* + \zeta (\zeta - 1)t_*$, eliminates the linear term from Richards' equation for $\Theta(Z, t)$:
\begin{align}
&\frac {\partial \Theta}{\partial t_*} =
\zeta^2 \frac {\partial }{\partial Z}\!
\left[
\frac {C(C-1)}{(C-\Theta)^2}
\frac {\partial \Theta}{\partial Z}
 \!-\! 
\frac {C(C-1)}{C-\Theta}
\right]
\end{align}

Introducing new independent and spatial variables
\begin{align}
&\omega(y, t_*) = 
\frac{(C-\Theta)}{\zeta(1-y)\sqrt{C(C-1)}},
\\
&y = 1- e^{-Z},\quad y_s(t_*) = 1- e^{-\zeta z_{s*}(t_*)-\zeta(\zeta-1)t_*},
\nonumber
\end{align}
then produces a well-known integrable nonlinear diffusion equation
\begin{equation}
\frac {\partial \omega}{\partial t_*} = 
\frac {\partial }{\partial y}
\left( 
\frac {1}{\omega^2}
\frac {\partial \omega}{\partial y}
\right)
\end{equation}
with associated boundary conditions
\begin{alignat}{2}
\omega\Big(y_s(t_*),t_*\Big) &=  \frac {\sqrt{C-1}}{\zeta\,(1-y_s(t_*))\sqrt{C}} \quad {\rm for} \quad &&t_* \in (0,\infty),
\\
\omega(y,0) &= \frac {\sqrt{C}}{\zeta\,(1-y)\sqrt{C-1}}
\, \quad {\rm for} \quad &&y \in (0,1).
\nonumber
\end{alignat}
From here we can proceed via a two-step process incorporating the reciprocal B\"acklund transformation \citep{Rogers1986IJNM}, or a two-step process using the hodograph transformation. Both are equivalent to introducing the new spatial independent variable 
\begin{align}
u = \int_{y_s(t_*)}^y \omega(y', t_*) \, dy' + u_s(t_*),
\end{align} 
and new dependent variable
\begin{align}
V(u,t_*) = \frac {\sqrt{C(C-1)}}{\zeta} \left[ \frac 1\omega - \frac {\zeta \sqrt{C-1}}{\sqrt{C}} (1-y) \right].
\end{align}
Utilising (\ref{fluxatfront}), to choose an appropriate $u_s(t_*)$ function:
\begin{align}
u_s(t_*) = \frac 1 {\sqrt{C(C-1)}} \Big[ 
i_*(t_*) + (\zeta(2C-1)-C) t_* + (C-1)z_{s*}(t_*)
\Big],
\label{usoft}
\end{align}
produces the linear heat equation with associated boundary conditions
\begin{alignat}{2}
\frac {\partial V}{\partial t_*} &= 
\frac {\partial^2 V}{\partial u^2}
\\
V(u,0) &= 0
&&\qquad u \in (0,\infty),
\nonumber
\\
V(u_s(t_*),t_*) &=  e^{-\zeta z_{s*}(t_*) - \zeta( \zeta-1) t_*}
&&\qquad t_* \in (0,\infty),
\nonumber
\\
V_u(u_s(t_*),t_*) &=  
-\frac {\sqrt C}{\sqrt{C-1}}
\Big( \zeta + i_*'(t_*) -1\Big)
e^{-\zeta z_{s*}(t_*) - \zeta( \zeta-1) t_*}
&&\qquad t_* \in (0,\infty).
\nonumber
\end{alignat}

The leading-order problem above has a Boltzmann scaling symmetry, and we adopt the canonical coordinates of this symmetry by choosing a new spatial variable $Y =  {u}/{\sqrt{t_*}}$, so the governing equation for $V(Y,t_*)$:
\begin{align}
t_* \frac {\partial V}{\partial t_*} = \frac Y2  \frac {\partial V}{\partial Y} + \frac {\partial^2 V}{\partial Y^2}.
\end{align}
is amenable to separation of variables, admitting Kummer confluent hypergeometric functions as solutions, see \citet{AbrSte1965}. The solution series 
\begin{equation}
V(Y,t_*) = e^{-Y^2/4} \sum_{n=0}^\infty C_n t_*^{\frac n2} 
\Psi\left(
\frac 12 + \frac n2, \frac 12, \frac {Y^2}4
\right),
\end{equation}
satisfies the our simple initial condition, with a sequence of separation constants $\{ C_j \}$ to be determined according to the Stephan boundary conditions
\begin{align}
V\left(\frac {u_s(t_*)}{\sqrt{t_*}},t_*\right ) &= e^{-\zeta z_{s*}(t_*) - \zeta( \zeta-1) t_*},
\label{finalBCs}
\\
V_Y\left(\frac {u_s(t_*)}{\sqrt{t_*}},t_*\right ) &= 
 -\frac {\sqrt C}{\sqrt{C-1}}
 \sqrt{t_*}
\Big( \zeta + i_*'(t_*) -1\Big)
e^{-\zeta z_{s*}(t_*) - \zeta( \zeta-1) t_*}.
\nonumber
\end{align}
Equation (\ref{usoft}) shows $u_{*s}(t_*)$ is a simple function of $i_*(t_*)$ and $z_{*s}(t_*)$. The closure condition (\ref{Darcy}) then gives $z_{*s}(t_*)$, as a function of the cumulative infiltration $i_*(t_*)$. It is convenient to represent this remaining unknown function as a power series in $\sqrt{t_*}$
\begin{equation}
i_*'(t_*) - 1 = \sum_{n=0}^\infty q_n t_*^{\frac {n-1}2}.
\label{infrateseries}
\end{equation}
Hence we have two sequences of unknown constants, the $\{ C_n \}$ and the $\{ q_n \}$, to be determined from the two boundary conditions (\ref{finalBCs}). Appendix \ref{AppA} describes efficient algorithms for determining these sequences. 

The truncated series give an exact solution of the linear heat equation, which is equivalent to an exact solution of the nonlinear Richards equation with zero truncation error. The boundary conditions vary with $N$, but if $t_*$ is not too large, the truncated quantities 
\begin{equation}
V(u,t_*) \simeq e^{-\frac {u^2}{4t_*}} \sum_{j=0}^N C_j t_*^{\frac j2} 
\Psi\left(
\frac 12 + \frac j2, \frac 12, \frac {u^2}{4t*}
\right), \quad 
i_*(t_*)  \simeq t_* + \sum_{n=0}^N \frac {2 q_n}{n+1} t_*^{\frac {n}2},
\end{equation}
are observed to satisfy the boundary conditions (\ref{finalBCs}) to greater accuracy as $N$ increases. Hence an error criterion can be adopted to select $N$ sufficiently large. The resulting solution $\Theta(z_*, t_*)$ in the unsaturated zone is given parametrically:
\begin{align}
&Z = \kappa \Big(u-u_s(t_*)\Big) - \ln 
\left[ e^{-\zeta z_{s*}(t_*) - \zeta( \zeta-1) t_*} - 
\frac {\zeta e^{-\kappa u_s(t_*)} } { \sqrt{C(C-1)}} 
\int_{u_c}^u e^{\kappa u' } V(u', t_*) \,du' \right],
\nonumber \\
&\kappa = \frac {\zeta \sqrt{C-1}}{\sqrt C}, \quad z_* = \frac Z\zeta - (\zeta-1) t_*, \quad 
\Theta = \frac {C V(u, t_*) e^Z}{C-1 + V(u, t_*) e^Z}.
\end{align}

\section{Soil moisture profiles}

\begin{figure}[h]
\begin{center}
\includegraphics[width=12cm]{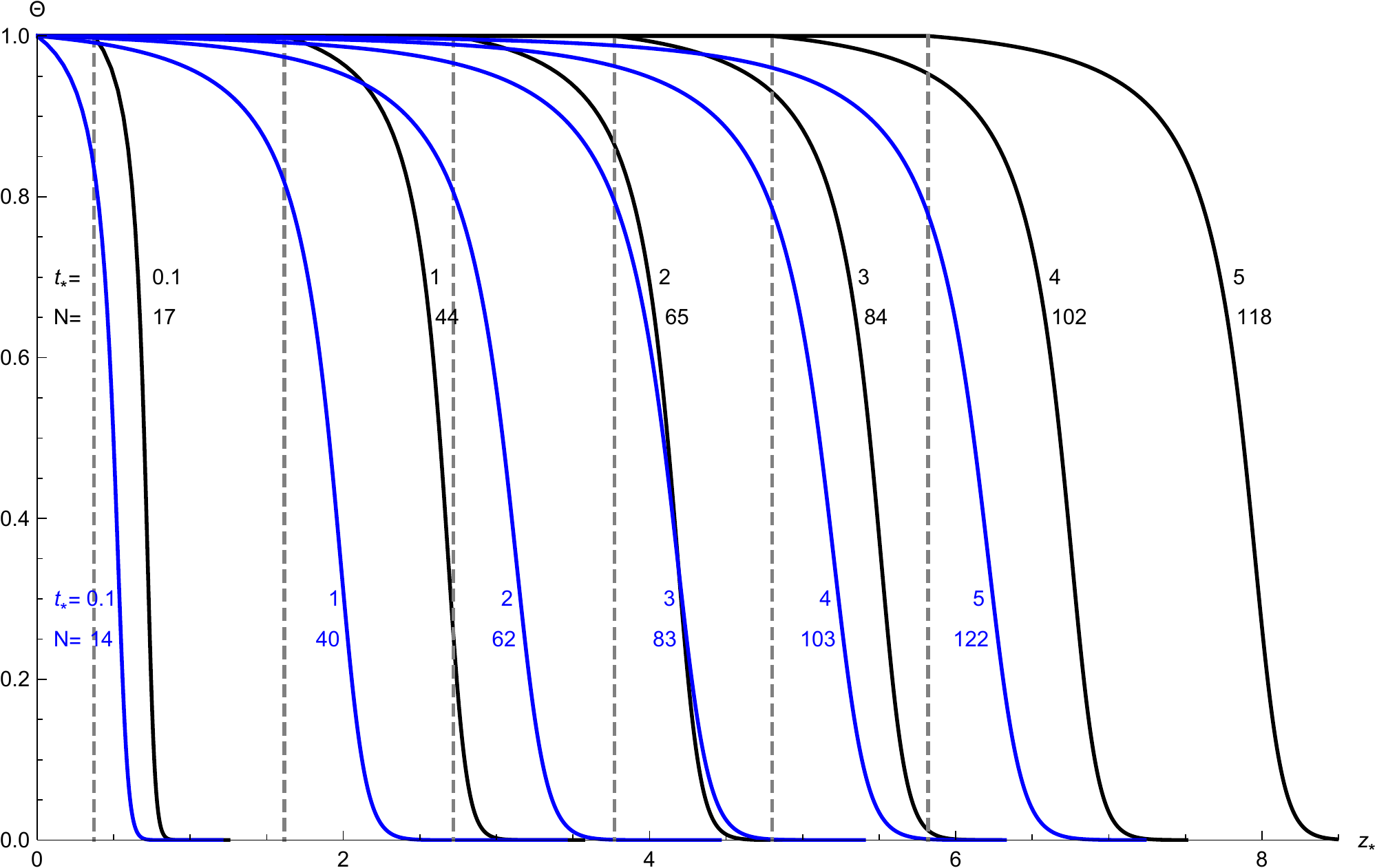}
\caption{Soil moisture content profiles for $C=1.1$, $\zeta = 1.05$. Results when subject to surface ponding with $h_+ = 1$ shown in black, and without surface ponding ($h_+= 0$) shown in blue.}
\label{h1wave}
\end{center}
\end{figure}
To plot soil moisture content profiles a suitable value of $N$ must be chosen. Our adopted criterion is for the ratio of the two sides of both our final boundary conditions (\ref{finalBCs}) to be within $1 \pm 10^{-6}$. Fig.~\ref{h1wave} shows results for a soil with $C=1.1$, $\zeta = 1.05$, either subject to surface ponding with $h_+ =1$, or subject to surface saturation without ponding with $h_+ =0$. As observed in \citet{TriadisBroadbridge2010} the series solution appears to converge at surprisingly large values of $t_*$, when soil moisture content evolution resembles large-time travelling wave behaviour. As time increases, more terms are required for the accurate satisfaction of boundary conditions, as shown by the displayed $N$ values. The dashed vertical lines show the position of $z_{*s}$ corresponding to each ponded soil moisture content profile. As expected, the increased surface pressure associated with surface ponding results in greater cumulative infiltration at all times. 

\begin{figure}[h]
\begin{center}
\includegraphics[width=12cm]{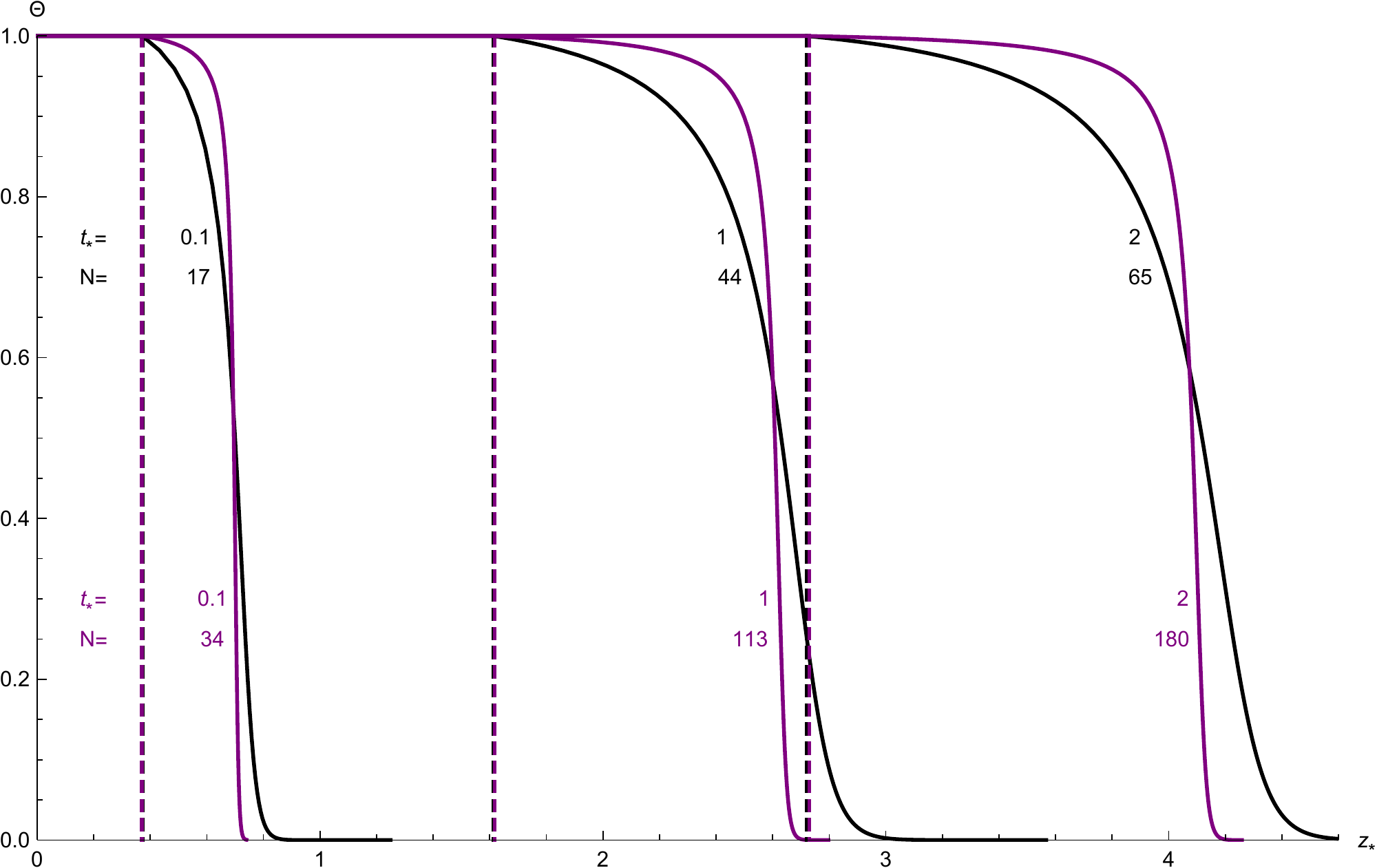}
\caption{Soil moisture content profiles for $h_+ =1$. Results for $C=1.1$, $\zeta = 1.05$ shown in black, compared with results for $C=1.02$, $\zeta = 1.01$ shown in purple.}
\label{h1waveCp02}
\end{center}
\end{figure}
We also consider the result when soils of differing properties are subject to the same level of surface ponding. Fig.~\ref{h1waveCp02} compares the results of the previous soil with $C=1.1$, $\zeta = 1.05$, with a soil that exhibits more sudden variation in hydraulic properties, such that $C=1.02$, $\zeta = 1.01$. Any change in the cumulative infiltration is not obvious from the soil moisture content curves, and only small changes in $z_{*s}$ are shown. The more nonlinear soil with a smaller value of $C$ requires significantly more terms in our series solution to satisfy (\ref{finalBCs}) to the same accuracy, and hence imposes a larger computational burden when calculating soil moisture content profiles.

\section{The delta-function-diffusivity limit}

As per the example in Fig.~\ref{h1waveCp02}, as $C$ decreases, the transition from $\theta_s$ to $\theta_n$ with increasing depth becomes steeper. In the mathematical limit as $C \to 1$ this transition occurs suddenly at some depth $z_c$ so that the soil is essentially saturated at lesser depths, and remains in its initial state $\theta_n$ at greater depths. This reduction of a smoothly-varying average soil moisture content to a step function is a popular modelling assumption, which includes the historically well-known \citet{GreenAmpt1911} model as a special case. In the current context the $C\to 1$ limit corresponds to $D(\theta)$ approaching a one-sided delta function positioned at $\theta_s$. See \citet{TriadisBroadbridge2010,TriadisBroadbridge2012} for further discussion. The present solution of Richards' equation allows this limit to be investigated analytically and unambiguously. However, some care must be taken to ensure the limit is approached in a realistic manner. A sensible choice is to hold the value of the sorptivity $S_0$ constant, in addition to other quantities $K_n$, $K_s$, $\theta_n$, $\theta_s$, $\gamma$, as done in \citet{TriadisBroadbridge2010}. Holding $S_0$ constant means that our adopted scales $l_s$ and $t_s$ are $C$-dependent. Consequently our dimensionless pond depth $h_+$ is $C$-dependent, and care must be taken to account for the limiting behaviour of the ratio $\mathfrak H/[C(C-1)]$ as $C\to 1$. When considering the nonlinear limit, we define an alternative dimensionless pond depth $\mathfrak h_+$, equal to the limiting value of $h_+$ as $C \to 1$:
\begin{equation}
h_+ = \frac {C(C-1)}{2 \mathfrak H} \mathfrak h_+, \quad \mathfrak h_+ = \frac {2(1+ \kappa_n)\Delta \theta\, \Delta K \,h}{S_0^2}.
\end{equation}
The cumulative infiltration $i_-(t)$ is the primary quantity of interest, and can be expressed as
\begin{align}
i_-(t) &= 2 q_0 \sqrt{ \frac {\mathfrak H}{C(C-1)}} S_0 \sqrt t + (q_1 + 1) \Delta K\, t 
\nonumber \\
& \quad + \sum_{n=2}^\infty \frac {2 q_n}{1+ n} \left( \frac{C(C-1)}{\mathfrak H} \right)^{\frac {n-1}2} \frac {(\Delta K)^n}{S_0^{n-1}} t^{\frac {n+1}2},
\nonumber \\
&= \sum_{n=0}^\infty S_{+n}(C,\mathfrak h_+) \frac {(\Delta K)^n}{S_0^{n-1}} t^{\frac {n+1}2}.
\end{align} 
The final form above serves to define dimensionless infiltration coefficients $\{ S_{+n} \}$
that are suitable for consideration of the $C\to 1$ limit. Through the careful use of software capable of symbolic manipulation, the following asymptotic behaviour can be determined
\begin{align}
S_{+0} &= \sqrt{1+{\mathfrak h_+}}
+\frac{{\mathfrak h_+}  (C-1)}{2 (1+{\mathfrak h_+})^{3/2}}
+\frac
{(C-1)^2}
{8 (1+{\mathfrak h_+})^{7/2}}
\left[-24 {\mathfrak h_+}-41 {\mathfrak h_+}^2-12 {\mathfrak h_+}^3 \right]
\nonumber \\ & \quad 
+O\left((C-1)^3\right),
\label{sp0eqn}
\\
S_{+1} &=
\frac{2-\zeta +2 {\mathfrak h_+}}{3 (1+{\mathfrak h_+})}
+ \frac 
{C-1}
{3 (1+{\mathfrak h_+})^3}
\left[-2+ 4 \zeta +   {\mathfrak h_+}(-2+ 7 \zeta  )\right]
\nonumber \\
&\quad + \frac
{(C-1)^2}
{3 (1+{\mathfrak h_+})^5} 
\big[14-28 \zeta  
+ {\mathfrak h_+} (34-86 \zeta)    
+ {\mathfrak h_+}^2 (18-64 \zeta)
\nonumber \\ &\quad 
+  {\mathfrak h_+}^3(-2+9\zeta)
+ 3 \zeta {\mathfrak h_+}^4
\big]
 +O\left((C-1)^3\right),
\label{sp1eqn}
\\
S_{+2} &= 
\frac
{1-\zeta+\zeta ^2 + {\mathfrak h_+}\left(2-\zeta +3 \zeta ^2\right) +{\mathfrak h_+}^2}
{9 (1+{\mathfrak h_+})^{5/2}}
\nonumber \\
& \quad +\frac
{(C-1)}
{18 (1+{\mathfrak h_+})^{9/2}}
\big[
-4+28 \zeta -28 \zeta ^2 
+{\mathfrak h_+}\left( -9  +97 \zeta -113 \zeta ^2  \right)
\nonumber \\ & \quad
+{\mathfrak h_+}^2 \left( -6 +69 \zeta -135 \zeta^2  \right)
+{\mathfrak h_+}^3\left(-1 -6 \zeta ^2 \right)
\big]
\nonumber \\ & \quad
+\frac
{(C-1)^2 }
{72 (1+{\mathfrak h_+})^{13/2}}
\big[336-1824 \zeta+1824 \zeta ^2
+{\mathfrak h_+} \left(1296 -8272 \zeta + 9104 \zeta ^2 \right) 
\nonumber \\ & \quad
+{\mathfrak h_+}^2\left(1611 -12531 \zeta +16587 \zeta^2\right) 
+{\mathfrak h_+}^3 \left(690  -6023 \zeta  +11205 \zeta ^2 \right) 
\nonumber \\ & \quad
+{\mathfrak h_+}^4 \left(51+84 \zeta +432 \zeta ^2\right) 
+{\mathfrak h_+}^5 \left(12 +24 \zeta  -48 \zeta ^2 \right) 
\big]
+O\left((C-1)^3\right).
\label{sp2eqn}
\end{align} 
The special case $\mathfrak h_+ =0$ corresponds to infiltration under surface saturation without ponding, and the above equations are in agreement with the corresponding terms of equations (B1) and (B2) of \citet{TriadisBroadbridge2010}. 

Limiting behaviour as the diffusivity approaches a delta function can also be determined by simplifying Richards' equation at the outset, rather than considering the limit of a more general exact solution as done above. \citet{Triadis2014} considered the more difficult case of falling head ponded infiltration, but equation (13) of that study can be solved for the limiting cumulative infiltration under constant head. For $0< \zeta < 1$ we obtain $i_*(t_*)$ in implicit parametric form:
\begin{align}
t_* &= \frac { \mathfrak h_+} { i_*' -1} - \mathfrak h_+ \ln\!\left(\! 1 + \frac 1{ i_*' -1}\right)
+ \frac 1 \zeta \ln\!\left(\! 1 + \frac \zeta{ i_*' -1}\right)
- \frac 1 {1-\zeta} \ln\!\left(\! 1 + \frac {1-\zeta}{ i_*' -1+\zeta}\right),
\nonumber \\
i_* &= \frac { \mathfrak h_+} { i_*' -1}+ \frac 1 \zeta \ln\!\left(\! 1 + \frac \zeta{ i_*' -1}\right).
\label{directC1eqns}
\end{align}
Substituting the series form of the $i_*'(t_*)$ (\ref{infrateseries}) into either of the these equations correctly reproduces the leading order terms of $S_{+0}$ to $S_{+2}$ already shown. We can also verify that substituting $\mathfrak h_+ = 0$ above, produces equation (7) of \citet{TriadisBroadbridge2012}, similar to equation (13) of \citet{ParlangeLisleBraddockSmith1982}. 

Given the limiting form of $K(\theta)$, the only physically sensible choice as $C \to 1$ in the nonlinear limit is to set $\zeta = 1$. This corresponds to a Gardner model soil and produces some simplification to equations (\ref{directC1eqns}):
\begin{align}
t_* &= \frac { \mathfrak h_+} { i_*' -1} - \frac {1} { i_*'} 
+(1-\mathfrak h_+) \ln\!\left(\! 1 + \frac 1{ i_*' -1}\right),
\nonumber \\
i_* &= \frac { \mathfrak h_+} { i_*' -1}+  \ln\!\left(\! 1 + \frac 1{ i_*' -1}\right).
\label{directC1eqnsz1}
\end{align} 
These are known equations, and occur in the delta-function-diffusivity limit of the models of \citet{ParlangeHaverkampTouma1985} and \citet{HaverkampParlangeStarrSchmitzFuentes1990}. They are almost identical to equations (2) and (5) of \citet{Parlange1975b}, which involve an `effective' hydraulic conductivity $K_{e}$. Parlange's equations become identical to those above if $H_0$ is first replaced by $h_0 K_s/K_{e}$, then remaining occurrences of $K_{e}$ replaced by $\Delta K$. 
Setting $\mathfrak h_+ = 0$ reproduces equation (8) of \citet{Parlange1975b}, also specified implicitly in equation (7) of \citet{TalsmaParlange1972}: 
\begin{equation}
t_* = i_* + e^{-i_*} -1.
\end{equation}

Finally, considering $\zeta = 0$ in the delta-function-diffusivity limit of our integrable soil results in an unphysical linear conductivity function, thereby overestimating conductive effects. Equations (\ref{directC1eqns}) are thus reduced to 
\begin{equation}
\frac {t_*}{1 + \mathfrak h_+} = \frac {i_*}{1 + \mathfrak h_+} -
\ln\!\left(\! 1 + \frac { i_*}{ 1 + \mathfrak h_+}\right),
\end{equation}
the familiar implicit form of the \citet{GreenAmpt1911} infiltration function with the ponded depth incorporated as a simple variable rescaling. Unfortunately this gives an unphysically high value of the first infiltration coefficient $ \frac 23 K_s$ in the limit of zero pond depth and $K_n=0$ \citep{TriadisBroadbridge2012}. \\

If we consider the behaviour of the leading-order terms of our infiltration coefficients (\ref{sp0eqn})--(\ref{sp2eqn}) as $\mathfrak h_+ \to \infty$, we find
\begin{align} 
S_{+0} &= \sqrt{ \mathfrak h_+} + O\big(\mathfrak h_+^{-1/2}\big) + O(C-1),
 \nonumber \\
S_{+1} &= \frac 23 + O\big(\mathfrak h_+^{-1}\big) + O(C-1),
 \nonumber \\
S_{+2} &= \frac 1{9\sqrt{\mathfrak h_+}} + O\big(\mathfrak h_+^{-3/2}\big) + O(C-1),
 \nonumber \\
S_{+3} &= -\frac 4{135 \mathfrak h_+} + O\big(\mathfrak h_+^{-2}\big) + O(C-1).
\end{align}
Here the behaviour of $S_{+3}$ has also been included, and we see an emerging trend, $S_{+n} = O\big(\mathfrak h_+^{-(n-1)/2}\big) + O(C-1)$, which implies an increased temporal range of convergence of our series solution for large values of $\mathfrak h_+$. In fact, for $C=1.1$, $\zeta = 1.05$, $h_+ = 1$ and $t_* \simeq 6.5$, we observe that boundary condition accuracy does not steadily increase for larger values of $N$. For $C=1.1$, $\zeta = 1.05$ and $h_+ = 10$ these impediments to boundary condition satisfaction are not observed until $t_* \simeq 27$. 

\section{Comparison with approximate models}

The ponded infiltration models of \citet{ParlangeHaverkampTouma1985} and \citet{HaverkampParlangeStarrSchmitzFuentes1990} are applicable to a wide range of soil properties, but have various approximations in their derivation. In contrast, the present model is an exact solution of the Richards equation, but applies only to an integrable model soil with properties as in equation (\ref{intmodelDK}). Hence the current model provides a test of the performance of the more general, approximate theory. 

\subsection{The infiltration function of \citet{ParlangeHaverkampTouma1985}}

Here cumulative infiltration subject to conditions of surface ponding $i_*(t_*)$ is specified according to two dimensionless parameters $\delta$ and $\gamma$. From equation (1) of \citet{ParlangeHaverkampTouma1985} $\delta$ can be stated in general using our present dimensionless variables, and evaluated specifically for the integrable model soil 
\begin{align}
\delta &= \int_0^1 \big(1-K_+(\Theta)\big) d\Theta,
\nonumber \\
& = \frac {1 + 2 \zeta C - \zeta}{2} - \zeta C (C-1) \ln\left( \frac C{C-1}\right).
\end{align}
The second parameter $\gamma$ incorporates the height of surface ponding and is defined in equations (19b) and (25) of \citet{ParlangeHaverkampTouma1985} 
\begin{equation}
\gamma = \frac
{\mathfrak h_+\left( 1 + \mathfrak h_+ + \frac {\mathfrak H}{C(C-1)} \int_0^1 (1-\Theta) D_*(\Theta) d\Theta \right) }
{1 + \mathfrak h_+ + \mathfrak h_+\left( 1 + \mathfrak h_+ + \frac {\mathfrak H}{C(C-1)} \int_0^1 (1-\Theta) D_*(\Theta) d\Theta \right) }.
\end{equation}
For the present soil properties we have 
\begin{equation}
\int_0^1 (1-\Theta) D_*(\Theta) d\Theta = (C-1) \left[ \ln\left( \frac C{C-1} \right) -1 \right].
\end{equation}
The infiltration $i_*(t_*)$ is then given in parametric form in equations (26) and (28) of \citet{ParlangeHaverkampTouma1985}
\begin{align}
\frac {2 \mathfrak H}{C(C-1)} t_* &= \frac 1 { \delta(1-\delta) } \ln \left( 1 + \frac \delta{i_*' -1} \right) + \frac \gamma {1-\gamma} \frac 1 {i_*' -1}
\nonumber \\ &\quad 
- \frac { 1- \gamma \delta} {(1-\gamma)( 1- \delta) } \ln \left( 1 + \frac 1{i_*' -1} \right),
\nonumber \\
\frac {2 \mathfrak H}{C(C-1)} i_* &= \frac \gamma{ 1- \gamma} \frac 1{i_*' -1} + \frac 1 \delta \ln \left( 1 + \frac \delta{i_*' -1} \right).
\label{PHT1985eqn}
\end{align}
Infiltration coefficients $S_{+n}$ can be derived given the above implicit specification of $i_*(t_*)$. As indicated by the $C$-dependence of $\delta$ and $\gamma$ shown above, the asymptotic forms of these infiltration coefficients have a different structure in the limit as $C\to1$, with first order corrections of order $(C-1) \ln (C-1)$, rather than $O(C-1)$. 

For $\mathfrak h_+ = 0$, $\gamma = 0$, and equations (\ref{PHT1985eqn}) reduce to the single implicit equation
\begin{equation}
(\delta - 1) \frac {2 \mathfrak H}{C(C-1)} t_* = \ln \left( \frac 1 \delta \left[ \exp \left( \delta \frac {2 \mathfrak H}{C(C-1)} i_* \right) + \delta -1 \right] \right) -  \frac {2 \mathfrak H}{C(C-1)} i_*, 
\label{PHT1985eqnlimit}
\end{equation}
equivalent to equation (13) of \citet{ParlangeLisleBraddockSmith1982}. 

\subsection{The infiltration function of \citet{HaverkampParlangeStarrSchmitzFuentes1990}}

\begin{figure}[]
\begin{center}
\includegraphics[width=12cm]{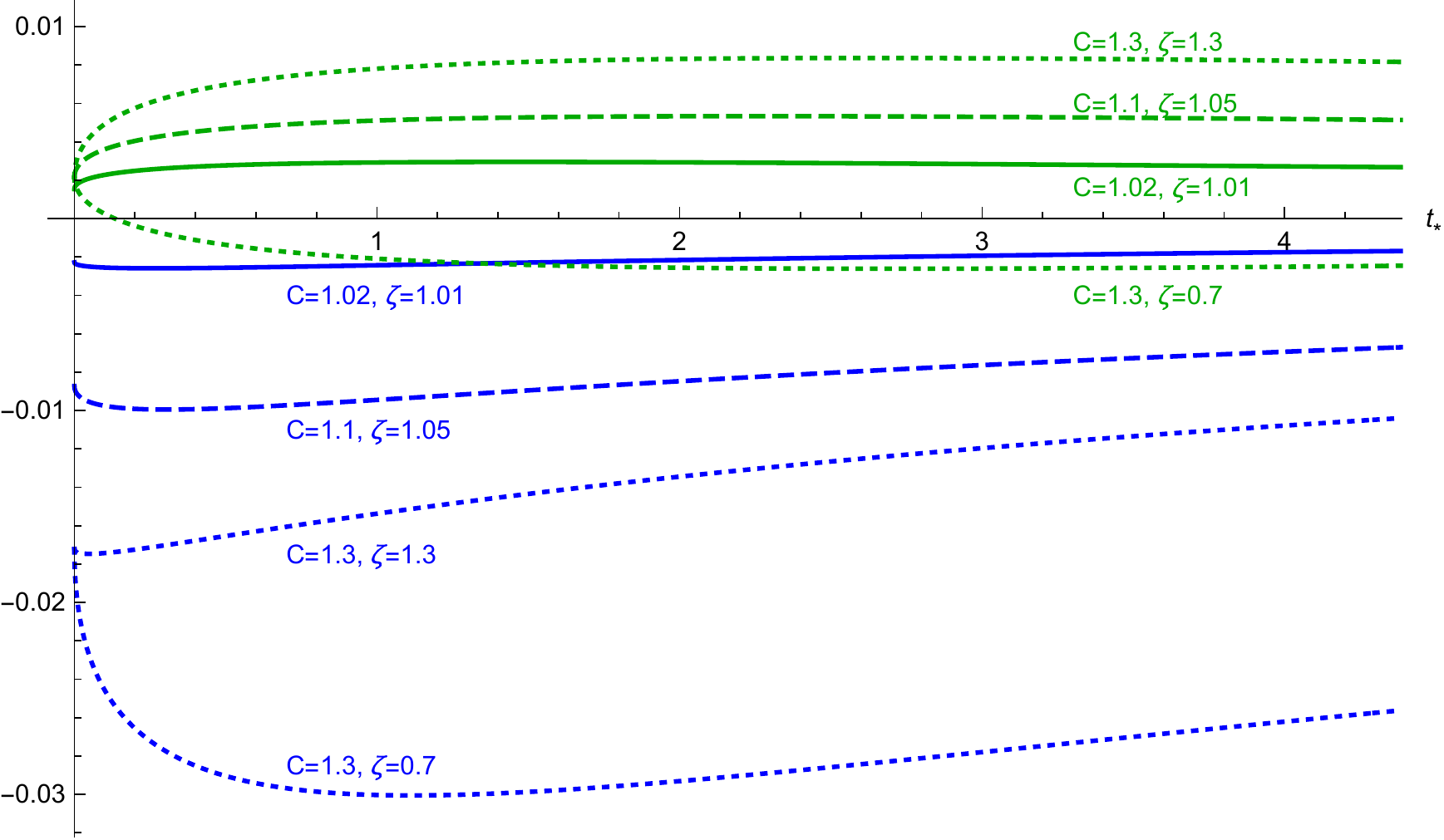}
\caption{Relative differences $i_{*m}(t_*)/i_*(t_*) -1$ for $h_+ =1$ and various soil properties. Results for the model of \citet{ParlangeHaverkampTouma1985} shown in green are compared with results for the model of \citet{HaverkampParlangeStarrSchmitzFuentes1990} shown in Blue.}
\label{Parlange1}
\end{center}
\end{figure}
This infiltration model shown in equations (22) and (23) of \citet{HaverkampParlangeStarrSchmitzFuentes1990} is similar in structure to that of \citet{ParlangeHaverkampTouma1985}, however the parameter $\delta$ has been set to $1$, and a new parameter $h_{str}$ is introduced, corresponding to an infinitely steep portion of the soil moisture potential curve. In equation (20) of \citet{HaverkampParlangeStarrSchmitzFuentes1990} the parameter $\gamma$ is redefined to incorporate $h_{str}$, but also simplified by eliminating dependence on the diffusivity
\begin{equation}
\gamma = \frac { \mathfrak h_+ - \mathfrak h_{str} } { 1 + \mathfrak h_+ }, \quad {\rm where} \quad 
\mathfrak h_{str} = \frac {2(1+ \kappa_n)\Delta \theta\, \Delta K \,h_{str}}{S_0^2}.
\end{equation}
The resulting two-parameter model is more clearly expressed in terms of $\mathfrak h_+$ and $\mathfrak h_{str} $ 
\begin{align}
\frac {2 \mathfrak H}{C(C-1)} t_* &=  \frac {\mathfrak h_+ - \mathfrak h_{str}}{i_*' -1} - \frac {1 +  \mathfrak h_{str}}{i_*'}
+ (1- \mathfrak h_+ + 2 \mathfrak h_{str} )  \ln \left( 1 + \frac 1{i_*' -1} \right)  ,
\nonumber \\ 
\frac {2 \mathfrak H}{C(C-1)} i_* &= \frac {\mathfrak h_+ - \mathfrak h_{str}}{i_*' -1} + (1 +  \mathfrak h_{str} )  \ln \left( 1 + \frac 1{i_*' -1} \right).
\label{HPSSF1990eqn}
\end{align}
In the present setting where our soil properties are specified by (\ref{intmodelDK}), $h_{str} = 0$, and the above model is almost identical to that shown in equation (\ref{directC1eqnsz1}) for the $C\to1$, $\zeta \to 1$ limit of the integrable soil. 

For $\mathfrak h_+ = 0$, equations (\ref{HPSSF1990eqn}) reduce to the single implicit equation
\begin{equation}
\frac {2 \mathfrak H}{C(C-1)} t_* = \frac {2 \mathfrak H}{C(C-1)} i_* + \exp \left( -\frac {2 \mathfrak H}{C(C-1)} i_* \right) - 1,
\label{HPSSF1990eqnlimit}
\end{equation}
shown as (8) of \citet{Parlange1975b}, and specified implicitly in (7) of \citet{TalsmaParlange1972}.
\medskip

\begin{figure}[]
\begin{center}
\includegraphics[width=12cm]{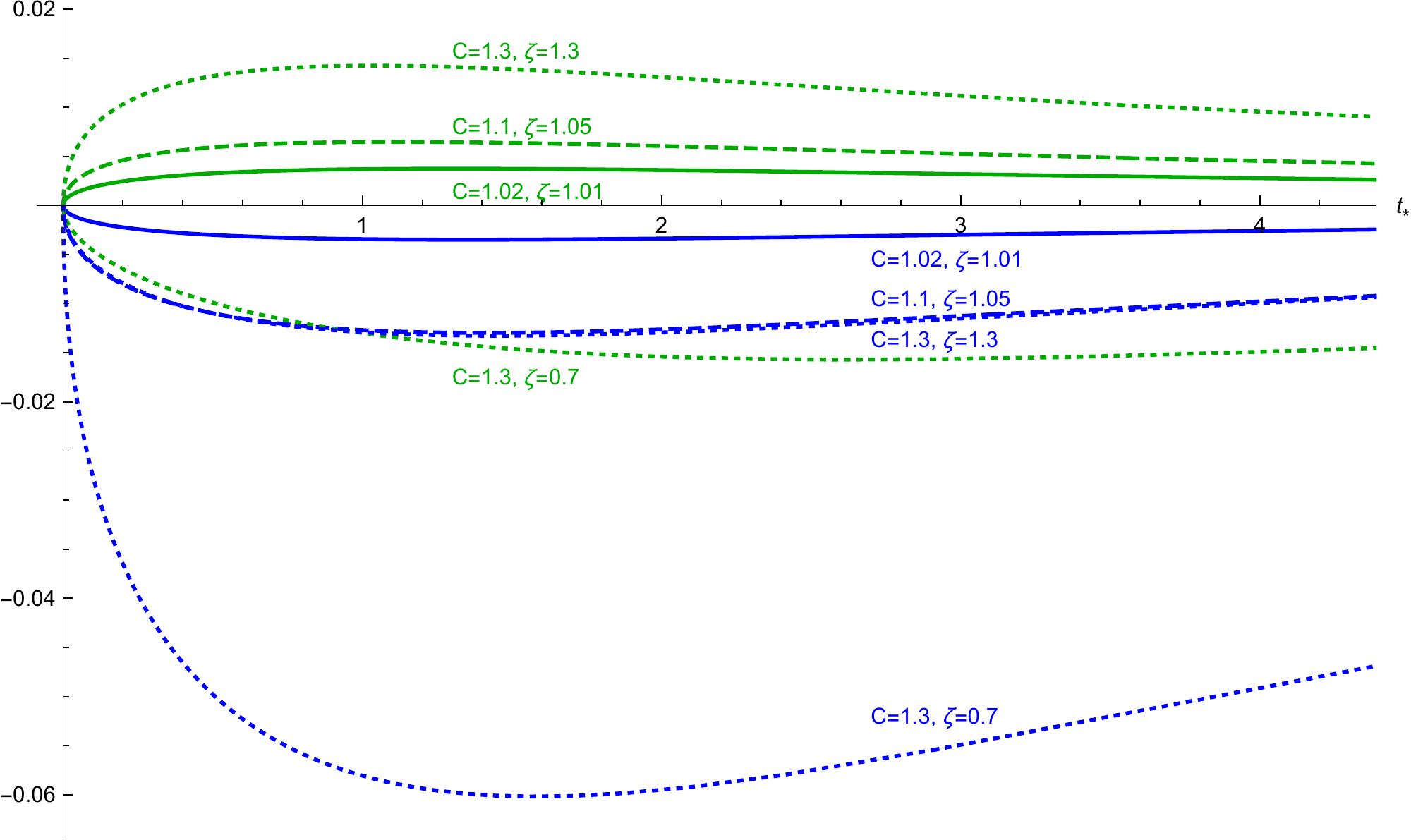}
\caption{Relative differences $i_{*m}(t_*)/i_*(t_*) -1$ for $h_+ =0$ and various soil properties. Results for the model of \citet{ParlangeHaverkampTouma1985} shown in green are compared with results for the model of \citet{HaverkampParlangeStarrSchmitzFuentes1990} shown in Blue.}
\label{Parlange0}
\end{center}
\end{figure}
If the cumulative infiltration according to one of the approximate models above is denoted $i_{*m}(t_*)$, Fig.~\ref{Parlange1} shows the relative difference $i_{*m}(t_*)/i_*(t_*) -1$ for $h_+ = 1$ and various soil properties. For the limited soil properties and ponding depths considered, both approximate models perform well relative to the exact solution, even for weakly nonlinear soils with $C=1.3$. For larger values of $C$ the model of \citet{ParlangeHaverkampTouma1985} appears to outperform that of \citet{HaverkampParlangeStarrSchmitzFuentes1990}. As $h_+$ increases, on the whole the performance of the approximate models considered improves, so that infiltration from saturated surface conditions without ponding, $h_+ = 0$, appears to be a worst-case scenario. Fig.~\ref{Parlange0} shows that even with $h_+ = 0$, the approximate models whose reduced forms are shown in equations (\ref{PHT1985eqnlimit}) and (\ref{HPSSF1990eqnlimit}), still perform very well. 

\section{Discussion and conclusion}

We have derived an exact solution of the nonlinear Richards equation subject to constant surface ponding, for integrable soil properties that are flexible enough to constitute realistic models for a class of field soils. The present study is a generalisation of a similar solution for surface saturation \citep{TriadisBroadbridge2010}, but now with the additional complication of a free boundary below the surface.  

The solution is in the form of a series in $\sqrt t$. Provided $t$ is in the domain of convergence of the series, the boundary conditions are achieved accurately after a sufficient number of terms are added. Due to the complex form of the solution, no precise results are known about the domain of convergence of the solution series for particular boundary conditions and soil properties. Nonetheless, the steady approach to very accurate satisfaction of boundary conditions can be computationally observed for a range of particular cases, enabling us to derive a useful range of soil moisture content profiles that exhibit travelling-wave-type behaviour for larger times. The integrable soil model and adopted solution technique also produce a family of exact, closed form solutions of the Richards equation that satisfy different boundary conditions, and the precise boundary conditions satisfied by the truncated solutions we have presented differ by a practically insignificant amount from constant surface ponding. 

The well-known \citet{GreenAmpt1911} model is just one example of a delta-function-diffusivity soil exhibiting simple plug-flow behaviour. The present integrable soil model allows behaviour in the delta-function-diffusivity limit to be convincingly determined within the framework of exact solutions to Richards' equation as the soil property $C$ approaches $1$. The present solution enables calculation of higher-order contributions proportional to $C-1$ and $(C-1)^2$ in addition to just leading-order terms, and a good number of easily implemented ponding depth-dependent infiltration coefficients have been provided in exact form. These contributions are confirmed to be compatible with less general, earlier solutions. As discussed in previous studies, in this framework the Green-Ampt model corresponds to an unphysical overestimation of conductive effects, hence we do not recommend it for practical soil modelling.

Some soil moisture content profiles presented require a truncated series length of nearly $200$ terms before boundary conditions are accurately satisfied, and for larger surface ponding depths $500$ terms may be needed for similar accuracy. Deriving hundreds of terms in our series solution is practically impossible without efficient, iterative algorithms to manage computational expense. The present study relies heavily on the groundwork established in \citet{TriadisBroadbridge2010}, with some generalisation required. Derivation of drastically simplified asymptotic behaviour in the $C\to1$ limit is also computationally taxing, and requires mathematical software capable of calculations incorporating very large symbolic expressions.  

We have shown that the exact solution constitutes a useful class of nontrivial test cases, for the validation of more versatile but approximate models of infiltration under surface ponding. For the various test cases considered, the impressive accuracy of the model of \citep{ParlangeHaverkampTouma1985} is a pleasing observation. In addition to increasing our fundamental knowledge of infiltration, our solution also provides nontrivial test cases for computational methods. 

Future work will consider variable head ponded infiltration, where the depth of surface water varies dynamically as a consequence of factors such as runoff, applied rainfall, and the cumulative infiltration itself. One important macroscopic prediction of such a generalised model is the time at which the ponded surface water is exhausted. It is unknown whether the domain of converge of the resulting series solutions will be sufficient to yield usable estimates of this characteristic time when $h$ falls to zero.  

\appendix 
\section{Satisfaction of transformed boundary conditions}
\label{AppA}

The remaining Stefan boundary conditions (\ref{finalBCs}) can be satisfied iteratively, by considering them order-by-order for small $t_*$. We will see that the leading-order produces transcendental equations which determine $C_0$ and $q_0$. For higher order expansions we find that the $n$-th order expansion of (\ref{finalBCs}) is linear in the $C_n$ and $q_n$, leading two a $2 \times 2$ linear system that is easily solved. 

\subsection{Leading-order equations}

Considering just leading-order satisfaction of our BCs specifies $q_0$ implicity by eliminating occurences of $C_0$:
\begin{align}
\gamma_0 &= \frac 1{\sqrt{C(C-1)}}\left( 2 q_0 + \frac {(C-1)h_+}{q_0}\right),
\nonumber \\
1 &= \sqrt{ \frac {\pi C}{C-1}} q_0 \,{\rm erfc}\left( \frac {\gamma_0}2 \right) \exp\!\left( \frac {\gamma_0^2}{4}\right).
\label{zeroeqn1}
\end{align}
This agrees with the calculation of \citet{Broadbridge1990}, which neglects gravitational effects.
The quantity $q_0$ in this study is equal to 
$\sqrt{C(C-1)/(4 h)}$ in the notation of \citet{Broadbridge1990}, and $h_+ = \Psi_{*0} $. If we assume the absence of a tension saturated zone so that $\Psi_{*s} = 0$, (\ref{zeroeqn1}) above is thus identified as equation (28) of \citet{Broadbridge1990}, verifying that the derived sorptivity under ponding is identical. 

Once $q_0$ and $\gamma_0$ are determined as above, the constant $C_0$ is given as 
\begin{align}
\frac 1 {C_0} = \sqrt \pi \,{\rm erfc} \left( \frac {\gamma_0}2 \right).
\end{align}

\subsection{Satisfaction at arbitrary order}

Expressing our boundary conditions in a form that isolates different higher orders in $\sqrt t$ involves rearranging hypergeometric series with unknown power series as arguments - this naively leads to considering sums over partitions of an integer $n$, which quickly become computationally infeasible \citep{BroadbridgeTriadisHill2009}. These partition sums were circumvented in \citet{TriadisBroadbridge2010} by using iterative methods to determine crucial expressions. We make use of these algorithms in the following description of a systematic method for determination of the constants $C_n$ and $q_n$, matching an arbitrary order $n$ in $\sqrt t$. 

Consider the following expansion for integer $j$
\begin{equation}
\left( \sum_{n=0}^\infty q_n t^{n/2} \right)^j = \sum_{n=0}^\infty t^{n/2} W_n(j; \{q_0, \ldots, q_n\}) = \sum_{n=0}^\infty x^n W_n(j; \{q_n\}).
\end{equation}
Here $W_n(j; \{q_n\})$ is just a condensed notation for the coefficient of $t^{n/2}$ when the power series is rearranged. This coefficient depends on the set of original coefficients $\{q_0, \ldots, q_n\}$, and is efficiently calculated iteratively as shown above equation (18) of \citet{TriadisBroadbridge2012} for $j<0$, and equation (A2) of \citet{TriadisBroadbridge2010} for $j>0$. 

Using (\ref{usoft}) and (\ref{infrateseries}) we have the power series form
\begin{align}
\frac {u_s(t_*)}{\sqrt{t}} &= \sum_{n=0}^\infty \gamma_n t_*^{n/2}, \quad \rm{where}
\\
\gamma_n &= \frac 1{\sqrt{C(C-1)}} \bigg( \frac {2 q_n}{1+ n} + (C-1)h_+ W_n(-1;\{q_n\}) 
\nonumber \\
&\quad+ \delta_{n1}[ \zeta(2C-1)-C+1]\bigg),
\nonumber \\
& = \frac {q_n}{\sqrt{C(C-1)}}\left(\frac 2 {n+1} - \frac {(C-1)h_+}{q_0^2}\right) + \gamma_{r,n};
\nonumber \\
\gamma_{r,n} &= \frac 1{\sqrt{C(C-1)}} \bigg( 
 \delta_{n1}[ \zeta(2C-1)-C+1] 
 \nonumber \\
 & \quad
 - \frac {(C-1)h_+}{q_0} \sum_{s=1}^{n-1} W_s(-1;\{q_s\}) q_{n-s} 
\bigg).
\nonumber
\end{align}
Here $\delta_{n1}$ is a Kronecker delta function. The final expression of $\gamma_n$ isolates occurrences of the highest order coefficient $q_n$ using the iterative relation for $W_n(-1; \{q_n\})$, and serves to define the remainder $\gamma_{r,n}$, which only contains coefficients $q_0, \ldots, q_{n-1}$ for $n> 0$.  

We also have the equivalence
\begin{align}
&e^{-\zeta z_{s*}(t_*) - \zeta( \zeta-1) t_*} = \sum_{n=0}^\infty Q_n t_*^{n/2}, \quad \rm{where}
\\
&Q_n = \sum_{i=0}^n \frac {\cos\left( \frac {(n-i)\pi}2 \right) \left[\zeta(\zeta-1)\right]^{\frac {n-i}2} }{\Gamma \left( 1 + \frac n2 - \frac i2\right) }
\sum_{m=0}^i \frac { (- \zeta h_+)^m }{m!} W_{i-m} \left( -m, ; \{ q_{i-m} \} \right).
\nonumber
\end{align}
Note that $Q_n$ is independent of $q_n$, containing only lower-order coefficients. 

Order-by-order expansion of of the LHSs of our final boundary conditions (\ref{finalBCs}) produces the novel series
\begin{align}
\xi(i, j) &= \sqrt \pi \sum_{m=0}^\infty
\frac
{(-1)^m W_i(m;\{\gamma_i\})}
{m! \,\Gamma(1 + \frac j2 - \frac m2)},
\\ 
\xi(0, j) &= 
e^{-\gamma_0^2/4} 
\Psi\left(
\frac 12 + \frac j2, \frac 12, \frac {\gamma_0^2}4
\right).
\nonumber
\end{align}
Note that in this work the $\xi$ will always relate to the set of $\gamma_i$ coefficients for $i = 0, 1, \ldots$, so we use the abbreviated notation $\xi(0, j)$, rather than $\xi(i, j;\{ \gamma_i \})$.  These are efficiently computed iteratively, according to equation (28) of \citet{Triadis2017} (note the typographical error in equation (A6) of \citet{TriadisBroadbridge2010}). This iterative relation also serves to define the dependence on the highest order coefficient, such that the remainder $\xi_r(i, j)$ can be isolated:
\begin{equation}
\xi(i, j) = \xi_r(i, j) - \gamma_i \xi(0, j-1).
\end{equation}
We adopt the following forms of the LHSs of (\ref{finalBCs})
\begin{align}
V\left(\frac {u_s(t_*)}{\sqrt{t_*}},t_*\right ) &= \sum_{l=0}^\infty t^{l/2} \sum_{m=0}^\infty C_{l-m} \xi(m, l-m),
\\
V_Y\left(\frac {u_s(t_*)}{\sqrt{t_*}},t_*\right ) &= - \sum_{l=0}^\infty t^{l/2} \sum_{m=0}^\infty C_{l-m} \xi(m, l-m-1).
\end{align}
Making use of the structure above, we have for $n > 0$, the $n$th order satisfaction of (\ref{finalBCs}) is equivalent to the following $2 \times 2$ linear system for $C_n$ and $q_n$
\begin{align}
&C_n \xi(0,n) + \frac {q_n C_0}{\sqrt{C(C-1)}} \xi(0,-1) \left( \frac {(C-1)h_+}{q_0^2} - \frac 2 {n+1} \right) = 
\nonumber \\
&\quad Q_n + C_0 \gamma_{r,n}  \xi(0,-1) - C_0 \xi_r(n,0) - \sum_{m=1}^{n-1} C_{n-m} \xi(m,n-m),
\\
&C_n \xi(0,n-1) + \frac {q_n}{\sqrt{C(C-1)}} \left( C_0 \xi(0,-2)\left[ \frac {(C-1) h_+}{q_0^2} - \frac 2 {n+1} \right] - C \right) = 
\nonumber \\
&\quad \sqrt{\frac C{C-1}}\left(\delta_{n1} \zeta + \sum_{m=0}^{n-1} Q_{n-m} (q_m + \delta_{m1}\zeta)\right) 
- \sum_{m=1}^{n-1} C_{n-m} \xi(m, n-m-1)
\nonumber \\
&\quad - C_0 \xi_r(n,-1) + C_0 \gamma_{r,n} \xi(0,-2). 
\end{align}
Hence the two coefficients $C_n$ and $q_n$ can be determined given known values of $C_0, \ldots, C_{n-1}$ and $q_0, \ldots, q_{n-1}$ . Evaluating the $\xi(i,j)$ and $W_n(j)$ quantities iteratively eliminates computational restrictions, so that sufficient coefficients can be evaluated to observe convergence or divergence of the series solution at particular times.

\bibliographystyle{dimdcu}
\bibliography{../../../Various/references}
\end{document}